\documentclass[trans,comsoc]{IEEEtran}
\usepackage{cite}
\usepackage{algorithm}
\usepackage{amsmath,amssymb,amsfonts}
\usepackage{algorithmic}
\usepackage{graphicx}
\usepackage{textcomp}
\usepackage{epstopdf}
\usepackage{stfloats}
\usepackage{setspace}
\ifCLASSOPTIONcompsoc
\usepackage[caption=false, font=normalsize, labelfont=sf, textfont=sf]{subfig}
\else
\usepackage[caption=false, font=footnotesize]{subfig}
\usepackage{bm}

\usepackage[T1]{fontenc}
\usepackage{cite}
\usepackage{amsmath,amssymb,amsfonts}
\usepackage{algorithmic}
\usepackage{graphicx}
\usepackage{textcomp}
\usepackage{xcolor}
\usepackage{bm}
\usepackage{changepage}
\usepackage{amsmath}
\usepackage{amstext}
\usepackage{stfloats}

\usepackage{amsmath}

\begin{document}

\title{\huge Learning Function-to-Function Mappings: A Fourier Neural Operator for Next-Generation MIMO Systems
	\thanks{
 \emph{Corresponding author: Ji Wang.}
	
Jian Xiao and Ji Wang are with the Department of Electronics and Information Engineering, College of Physical Science and Technology, Central China Normal University, Wuhan 430079, China (e-mail: jianx@mails.ccnu.edu.cn; jiwang@ccnu.edu.cn). 

Qi Sun and Chih-Lin I are with the China Mobile Research Institute, Beijing 100876, China. (e-mail: sunqiyjy@chinamobile.com; icl@chinamobile.com).

Qimei Cui is with National Engineering Research Center for Mobile Network Technologies, Beijing University of Posts and Telecommunications, Beijing 100876, China (e-mail: cuiqimei@bupt.edu.cn).
	
Xingwang Li is with the School of Physics and Electronic Information Engineering, Henan Polytechnic University, Jiaozuo 454003, China (e-mail: lixingwang@hpu.edu.cn).

Dusit Niyato is with the College of Computing and Data Science, Nanyang Technological University, Singapore 639798 (e-mails: dniyato@ntu.edu.sg).
	}
	}
\author{Jian Xiao, Ji Wang,~\IEEEmembership{Senior Member,~IEEE}, Qi Sun, Qimei Cui,~\IEEEmembership{Senior Member,~IEEE}, \\Xingwang Li,~\IEEEmembership{Senior Member,~IEEE}, Dusit Niyato,~\IEEEmembership{Fellow,~IEEE}, and Chih-Lin I,~\IEEEmembership{Life Fellow, IEEE}}
\maketitle
\begin{abstract}
Next-generation multiple-input multiple-output (MIMO) systems, characterized by extremely large-scale arrays, holographic surfaces, three-dimensional architectures, and flexible antennas, are poised to deliver unprecedented data rates, spectral efficiency and stability. However, these advancements introduce significant challenges for physical layer signal processing, stemming from complex near-field propagation, continuous aperture modeling, sub-wavelength antenna coupling effects, and dynamic channel conditions. Conventional model-based and deep learning approaches often struggle with the immense computational complexity and model inaccuracies inherent in these new regimes. This article proposes a Fourier neural operator (FNO) as a powerful and promising tool to address these challenges. The FNO learns function-to-function mappings between infinite-dimensional function spaces, making them exceptionally well-suited for modeling complex physical systems governed by partial differential equations based on electromagnetic wave propagation. We first present the fundamental principles of FNO, demonstrating its mesh-free nature and function-to-function ability to efficiently capture global dependencies in the Fourier domain. Furthermore, we explore a range of applications of FNO in physical-layer signal processing for next-generation MIMO systems. Representative case studies on channel modeling and estimation for novel MIMO architectures demonstrate the superior performance of FNO compared to state-of-the-art methods. Finally, we discuss open challenges and outline future research directions, positioning FNO as a promising technology for enabling the enormous potential of next-generation MIMO systems.

\end{abstract}


%
\IEEEpeerreviewmaketitle

\section{Introduction}
\IEEEPARstart{T}he relentless demand for higher data rates, lower latency, and massive connectivity is driving wireless communication systems towards sixth-generation (6G) networks. At the heart of this evolution lies the advancement of multiple-input multiple-output (MIMO) antennas. While current fifth-generation (5G) systems have successfully leveraged massive MIMO technology, the next generation of MIMO systems will push the boundaries of the physical layer, introducing new hardware paradigms and propagation characteristics that demand a fundamental rethinking of physical-layer signal processing techniques \cite{11026007}.

\subsection{Next-Generation MIMO systems}
\begin{figure*}[t]
\centering
\includegraphics[width=6.5in]{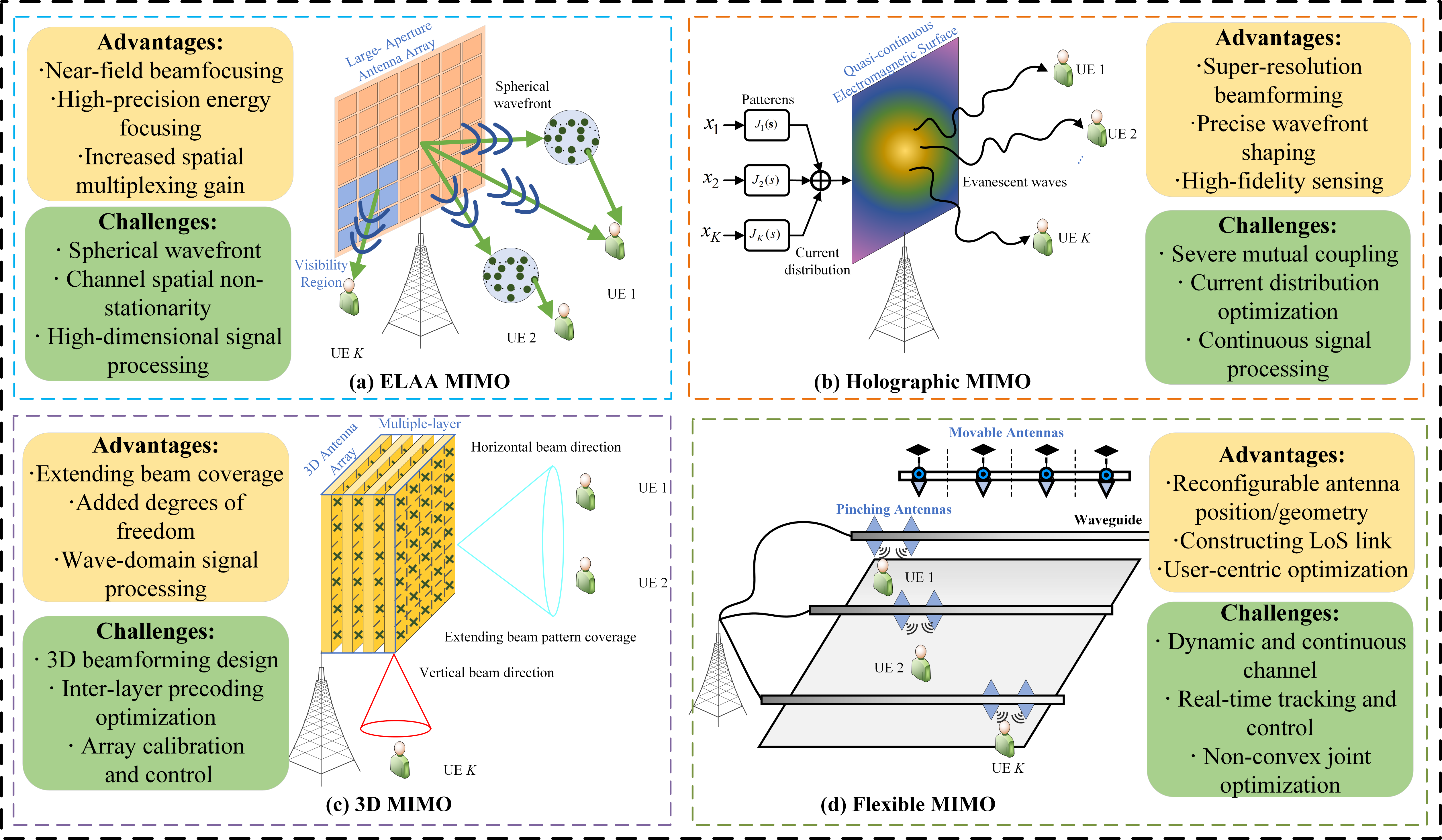}
\caption{Illustration of next-generation MIMO systems: (a) Extremely Large-Aperture Array (ELAA) MIMO, (b) Holographic MIMO, (c) 3D MIMO, and (d) Flexible MIMO.}
\label{fig1}
\end{figure*}

As illustrated in Fig.~\ref{fig1}, next-generation MIMO systems are expected to move beyond the conventional massive MIMO framework by incorporating several innovative architectures.

\subsubsection{Extremely Large-Aperture Array} ELAA, also known as ultra-massive MIMO \cite{akyildiz2016realizing}, will feature apertures of hundreds or even thousands of wavelengths by further integrating extremely large-scale antennas. At these scales, the electromagnetic (EM) propagation environment transitions from the far-field to the radiative near-field. In this regime, the conventional plane wave assumption breaks down, and spherical wavefronts must be considered. This leads to spatially non-stationary channel characteristics across the array, where the channel statistics vary depending on the antenna location. 

\subsubsection{Dense Antenna Array}
To enable the manipulation of the EM field with unprecedented resolution, the spacing between antenna elements can be reduced to a fraction of the wavelength \cite{10232975}, approaching the scale of holographic MIMO or continuous-aperture MIMO. However, it also introduces strong mutual coupling between antenna elements as antenna arrays become denser. Furthermore, dense arrays can capture evanescent waves, which decay exponentially with distance from the source but carry sub-wavelength information. Harnessing these effects can lead to superdirectivity and significantly enhanced spatial resolution, but requires sophisticated signal processing to manage the intricate interactions. 

\subsubsection{Three-Dimensional Antenna Array}
Moving from two-dimensional (2D) planar arrays to three-dimensional (3D) volumetric arrays \cite{10826975}, offers additional degrees of freedom for shaping the radio waves. 3D MIMO can provide full-space coverage and more resilient connections. In particular, a special case of 3D array, i.e., stacked intelligent metasurface \cite{11014597}, allows for processing in the wave domain of the propagating field. However, designing and controlling multiple-layer 3D arrays to overcome fundamental physical constraints, e.g., the Hannan limitation on directivity, present a formidable task.

\subsubsection{Flexible Antenna Array}
The concept of flexible antennas introduces a new dimension of adaptability. This includes systems with movable antennas, fluid antennas \cite{zhu2024historical}, and pinching-antennas that can change their positions \cite{Ding2024}, as well as flexible intelligent metasurfaces (FIMs) that can be conformally mounted on various surfaces \cite{11014597}. These systems can dynamically alter their physical geometry to optimize performance in response to changing environmental conditions or user locations. The challenge lies in the real-time modeling and control of the antenna positions and array geometry.

%

\subsection{Overarching Challenges and Motivations}
The push toward next-generation MIMO systems is fueled by demanding future network applications, which requires a significant increase in data rates, reliability, and spatial awareness. However, realizing this vision requires overcoming fundamental challenges at the wireless physical layer.

\subsubsection{The Breakdown of Conventional Assumptions}
Decades of wireless signal processing have been built upon simplified physical and statistical models that are rendered obsolete by next-generation MIMO systems. For ELAA systems, the shift to the radiative near-field invalidates the convenient far-field plane wave assumption. The spherical wavefronts induces spatial non-stationarity, where channel statistics vary across the array. This invalidates traditional channel models that assume spatial stationarity, making physical-layer signal processing tasks, such as channel estimation and precoding, significantly more difficult. For holographic MIMO systems, the ultra-dense packing of antennas introduces severe mutual coupling between adjacent antennas, which becomes a dominant factor that corrupts signal models.

\subsubsection{The Explosion in Complexity and Dimensionality}
The sheer scale of these new systems triggers an exponential growth in computational and signaling overhead. For ELAA and 3D MIMO systems, the number of antenna elements can grow to tens of thousands or more. This leads to channel matrices of enormous dimensions, making traditional signal processing techniques including matrix inversion for linear precoding computationally infeasible in real-time. This curse of dimensionality also dramatically increases the pilot overhead required for accurate channel estimation. For holographic MIMO systems, treating the aperture as continuous means moving from finite-dimensional matrices to infinite-dimensional functions and operators, a domain where traditional discrete signal processing algorithms are not applicable.

\subsubsection{The Demand for Extreme Adaptability}
The dynamic nature of future wireless applications, combined with new reconfigurable hardware, demands unprecedented levels of adaptability. For flexible MIMO systems, the challenge becomes more pronounced. The communication channel is dynamic not only because of external environmental factors, but also because the physical array geometry, e.g., the position and shape of its antennas, is actively and continuously changing.
The channel becomes a function of the physical state of the antenna array. This necessitates a new class of ultra-low-latency adaptive algorithms that can track these rapid, self-induced channel variations and co-optimize the physical configuration with the signal processing.

These intertwined challenges reveal the limitations of existing solutions. Traditional model-based signal processing algorithms are brittle, as their performance plummets when their underlying mathematical assumptions, e.g., sparsity and stationarity of wireless channels, are violated. Moreover, conventional neural networks struggle to scale to such high dimensions and, more critically, lack the ability to generalize to the continuous and dynamic variations in geometry inherent in these new systems. This creates an urgent need for a new signal processing paradigm that can learn the complex underlying physics, scale efficiently, and inherently adapt to dynamic functional spaces. 

\subsection{Contributions}
Against this background, this article proposes a Fourier neural operator (FNO) as a transformative tool for physical-layer processing in next-generation MIMO systems. The FNO is a novel class of neural operators that are mathematical constructs that implement mappings between function spaces \cite{lifourier}. This characteristic endows the FNO with an inherent advantage for modeling physical systems governed by partial differential equations (PDEs). In fact, wireless communication is essentially EM wave propagation, which can be described by Maxwell’s equations, constituting a core set of PDEs. In this EM propagation scenario, the wireless channel can be interpreted as the solution operator to these Maxwell’s equations. By leveraging the function-to-function mapping capability of the FNO, we reframe physical-layer signal processing problems in next-generation MIMO systems not as statistical learning tasks, but as learning the fundamental physics underlying the wireless channel.
\begin{itemize}
\item We first establish FNO as a learning framework that is structurally and philosophically aligned with the physics of wireless communications. We present the core principle of function-to-function mappings between infinite-dimensional function spaces. This allows FNO to overcome the resolution-dependency and model-mismatch limitations inherent in conventional deep learning and model-based methods.
\item We further develop a unified operator-theoretic framework that recasts a wide array of traditionally distinct physical layer tasks under the common paradigm of operator learning. This is achieved by systematically classifying each task based on the mathematical properties and functional role of the required operator, providing a structured methodology for applications of FNO.
\item Case studies and numerical results on holographic MIMO systems demonstrate that FNO can serve as a highly accurate surrogate for complex wave propagation physics. Furthermore, the FNO can significantly improve the dynamic channel estimation accuracy with less pilot overhead for FIM systems. Finally, open challenges and future directions are summarized for the deep application and deployment of FNO in next-generation MIMO systems.
\end{itemize}

\section{Principle of Fourier Neural Operator}
\begin{figure}[!t]
\centering
\includegraphics[width=3.0in]{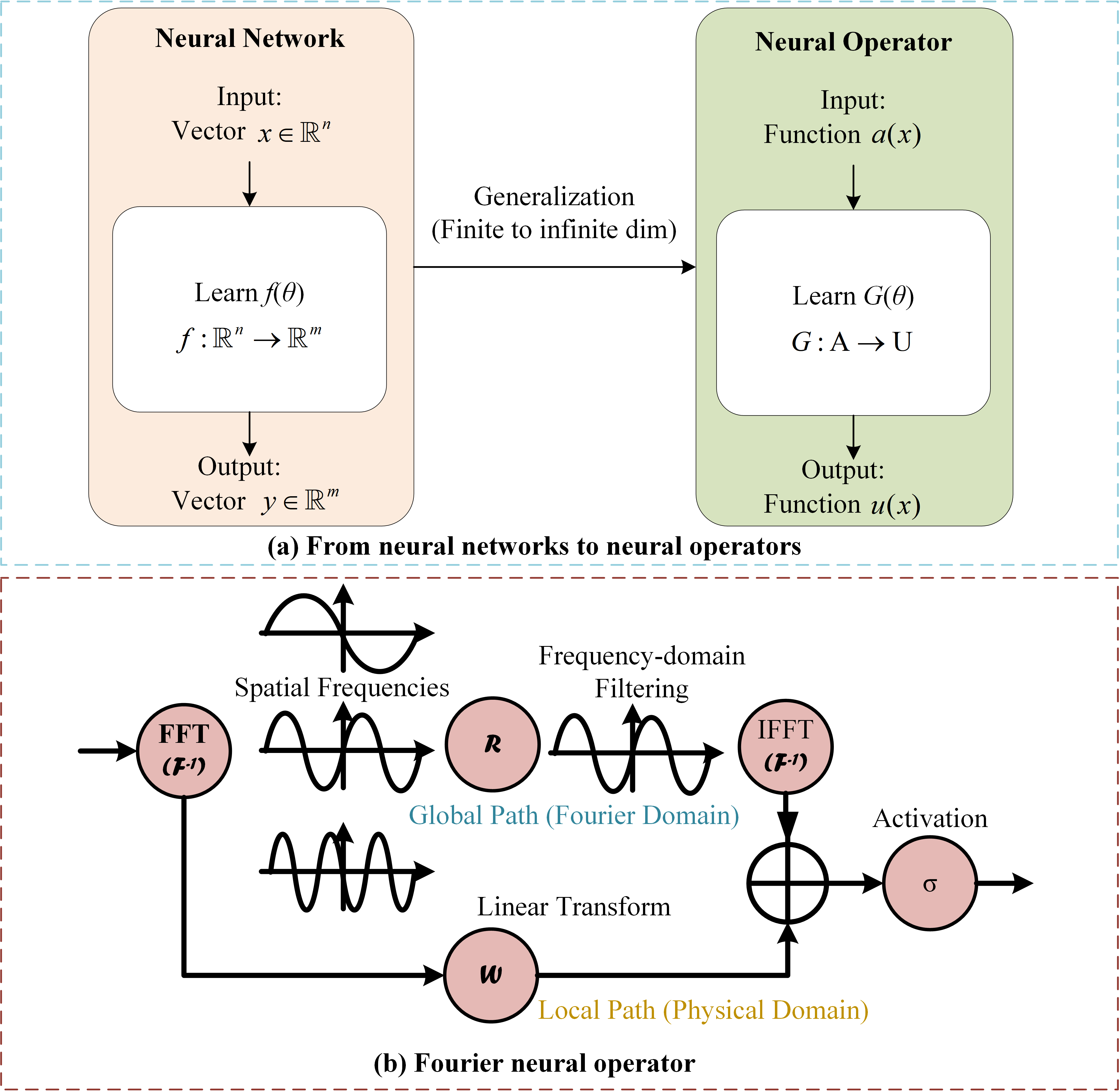}
\caption{Design principle of Fourier neural operator. (a) Evolution between neural networks and neural operators, and (b) Pipeline of Fourier neural operator.}
\label{fig2}
\end{figure}
In contrast to traditional neural networks that learn mappings between fixed-size vectors, FNO learns mappings between continuous functions, which is a fundamental shift in the learning paradigm.

\subsection{From Neural Networks to Neural Operators}

{As illustrated in Fig.~\ref{fig2}(a), a conventional deep neural network can be regarded as a powerful function approximator that learns a mapping from an input vector in a finite-dimensional space $\mathbb{R}^n$ to an output vector in $\mathbb{R}^m$ \cite{cui2025overview}. This results in a key limitation: such networks are dependent on the discretization of the data. However, many problems in science and engineering, including wireless communications, are more naturally described as interactions between functions defined over continuous domains. The wireless channel is a function of continuous variables, such as space, time, and frequency.}

{In this case, a neural operator can be designed to learn the underlying rule or physical law that maps an entire input function to an output function, i.e., a mapping between infinite-dimensional function spaces. For instance, it can learn the operator that maps a continuous function describing the 3D deformation shape of a flexible antenna array (an input function) to the corresponding continuous channel field distribution in space (an output function). This ability to work in infinite-dimensional function spaces allows neural operators to generalize beyond the specific discretization or resolution.}

\subsection{The Fourier Neural Operator Mechanism}
The FNO is a particularly efficient and effective type of neural operator. Its capability comes from performing the most complex part of its computation in the Fourier domain. Fig.~\ref{fig2}(b) presents a standard FNO layer that processes an input function through two parallel paths.

\begin{enumerate}
\item \textbf{Global path:} In this global Fourier path, the input function is first transformed into the frequency domain using the fast Fourier transform (FFT). In the Fourier domain, complex global convolution operations become simple element-wise multiplications, which are needed to capture long-range dependencies across the entire antenna array. The FNO applies a learnable filter to these frequency components, effectively learning which frequencies are important for the problem at hand. It then transforms the filtered signal back to the spatial domain using an inverse FFT. This path allows the FNO to learn the entire input function at once and efficiently capture global phenomena like wave propagation and reflection.
\item \textbf{Local path:} In parallel, a simple linear transformation is applied to the input function at each point individually. This path allows the FNO to capture local features and fine-grained details.
\end{enumerate}

The outputs of these two paths are then combined and passed through a non-linear activation function. By stacking several such layers, the FNO can learn extremely complex and non-linear operators, effectively modeling the intricate physics of the wireless channel.

\subsection{Suitability of FNO for Wireless Physical Layer}
\begin{figure}[!t]
\centering
\includegraphics[width=3.0in]{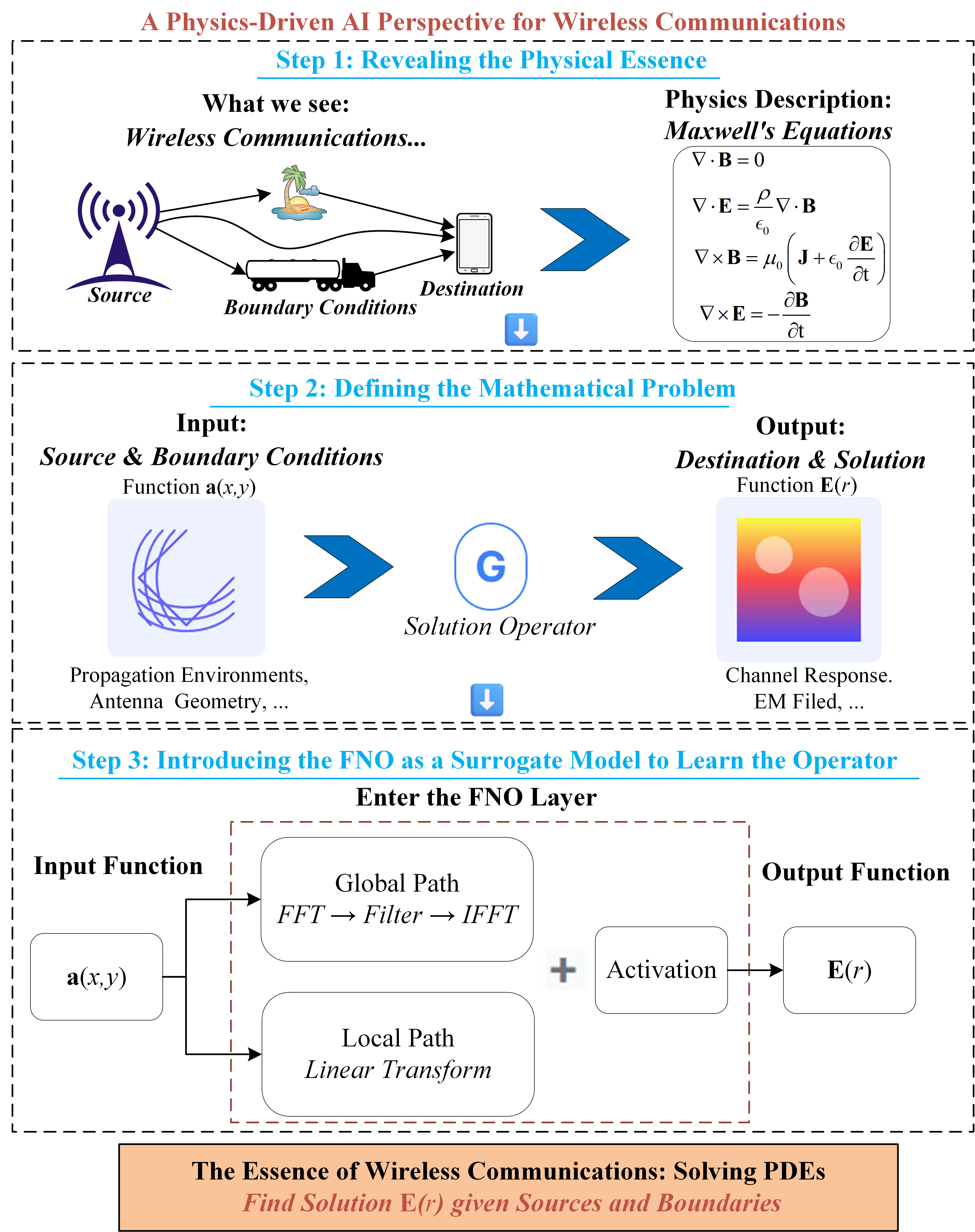}
\caption{{Connection between the physical essence of wireless communication and FNO. First, framing wireless communication as a physical process governed by Maxwell's equations; second, abstracting this process into a mathematical problem of finding a solution operator; and third, introducing the FNO as a surrogate model to learn this operator directly from data.}}
\label{Logic}
\end{figure}

As illustrated in Fig.~\ref{Logic}, the architecture of FNO inherently aligns with the physics of wireless communications, in which the capability of FNO stems from its design as a solver for systems governed by PDEs. This is profoundly relevant because the entire domain of wireless communication is fundamentally described by Maxwell's equations \cite{4020419}, i.e., a system of PDEs. The wireless channel itself, which dictates how signals propagate from a transmitter to a receiver, is a physical manifestation of the solution to these PDEs for a specific set of boundary conditions, e.g., the array geometry, environmental scatterers and reflectors. {The rule that maps these boundary conditions to the final field solution is known as the solution operator. The FNO is designed to be a universal approximator of this complex and non-linear solution operator. Furthermore, the reliance on Fourier transforms is not arbitrary. Many PDEs, including the wave equation, are most efficiently solved in the frequency domain, making FNO architecture structurally aligned with the underlying physics. The specific advantages of FNO are summarized as follows.


\begin{itemize}
\item \textbf{Learning continuous operators:} Wireless channels are fundamentally continuous functions of space, time, and frequency. FNO is designed to learn operators on these functions directly, rather than just memorizing discrete points. This allows it to estimate the channel response at any point on a holographic surface.
\item \textbf{Handling high-dimensional inputs:} Physical layer data is high-dimensional for next-generation MIMO systems. The utilization of global convolution in the Fourier domain is exceptionally well-suited for this. It can efficiently capture complex dependencies across all dimensions, such as how signals from multiple users interfere across a wide frequency band on a large antenna array.
\item \textbf{Parameter Efficiency:} A key feature of FNO is that its Fourier-domain filter typically focuses on a limited number of lower-frequency modes, as these often carry the most significant information. By truncating the high-frequency modes that corresponds to noise or negligible details, FNO can learn the underlying operator with significantly fewer parameters than those of conventional neural networks would need for the same task.
\item \textbf{Fusion with Physical Laws:} The structure of FNO resonates with the physics of electromagnetism. Maxwell's equations, which govern wave propagation, are often solved efficiently in the frequency domain. The reliance on Fourier transforms provides a natural bridge to embed physical priors or combine it with physics-based models.
\end{itemize}

\section{Applications of Fourier Neural Operator in Next-Generation MIMO Systems}
The FNO offers a unified perspective by reframing various physical-layer problems as specific types of operator learning tasks. Fig.~\ref{fig3} categorizes the roles FNO plays based on the mapping relationships it learns in different scenarios.
%

\begin{figure*}[t]
\centering
\includegraphics[width=6.5in]{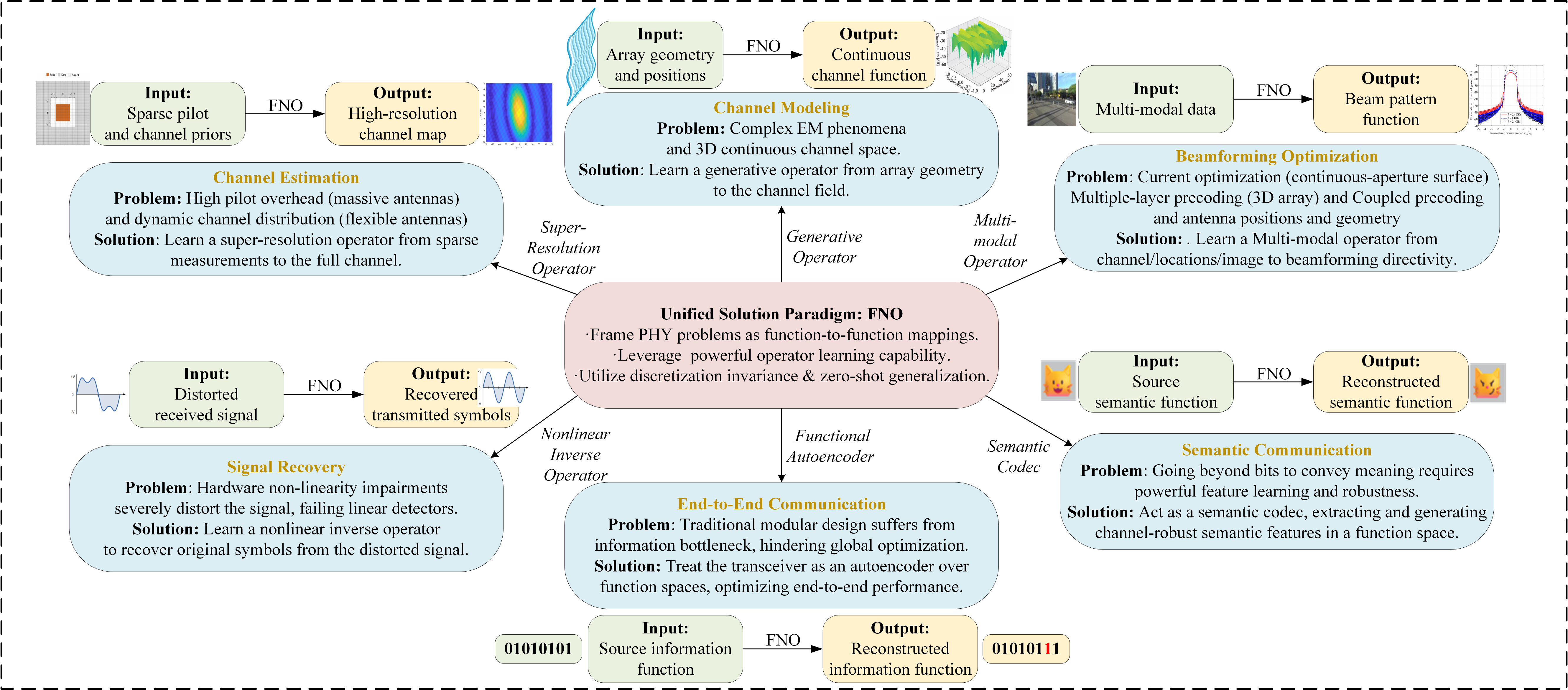}
\caption{Applications of FNO in physical-layer signal processing of next-generation MIMO systems. FNO provides a potential foundational tool that reframes diverse physical-layer challenges as specific function-to-function mapping tasks, where each branch illustrates a distinct application, detailing the problem, the role of the FNO as a learned operator and the corresponding inputs and outputs of the mapping task.}
\label{fig3}
\end{figure*}

\subsection{FNO as a Generative Operator}
\begin{itemize}
\item \textbf{Channel Modeling:} Accurate channel modeling is fundamental to next-generation MIMO system design. However, the complex EM phenomena in next-generation MIMO systems, e.g., near-field effects, spatial non-stationarity, continuous apertures, make analytical modeling exceedingly difficult, while traditional numerical simulations are computationally expensive \cite{gibson2021method}.

\item \textbf{FNO Solution:} FNO can act as a generative operator, learning the mapping from the physical environment and array geometry to the complete channel field distribution. For instance, the input could be a function describing the scatterer distribution and the geometric parameters of the array, with the output being the channel response at any point in space. This approach transforms FNO into an efficient physics-informed surrogate model, which can rapidly generate physically consistent channel data in the millisecond (ms) level per instance, thereby greatly accelerating system simulations and algorithm validation.
\end{itemize}

\subsection{FNO as a Super-Resolution Operator}
\begin{itemize}
\item \textbf{Channel Estimation: }  In next-generation MIMO systems, acquiring accurate channel state information (CSI) is complicate. Specifically, the large number of antennas results in prohibitive pilot overhead. This challenge is further amplified in Holographic and flexible MIMO systems with continuous and dynamic channels \cite{11018390}.

\item \textbf{FNO Solution:} FNO can be trained as a super-resolution operator that learns the mapping from sparse and low-resolution channel measurements to a complete and high-resolution channel map. This approach effectively addresses the near-field non-stationary effects in ELAA, the quasi-continuous apertures of holographic MIMO, as well as the geometry-dependent channels of flexible MIMO. The trained FNO excels at super-resolution channel estimation by generalizing across a continuum of physical configurations, which allows it to accurately map the entire channel  from only a limited number of pilot probes. 

\end{itemize}
\subsection{FNO as a Multi-Modal Operator}
\begin{itemize}
\item \textbf{Beamforming Optimization:} In holographic MIMO and 3D MIMO, beamforming is no longer about optimizing a discrete precoding vector but rather a continuous current distribution function or a complex multi-layer phase configuration \cite{10232975,10826975}. Furthermore, for flexible MIMO, the optimal beamforming is coupled with the physical positions of the antennas \cite{11014597}.

\item \textbf{FNO Solution:} To reduce the difficulty of non-convex beamforming, FNO can learn a multi-modal operator that maps diverse inputs, e.g., user locations, beam patterns, or even semantic information from images, to the desired beamforming configuration. This allows for CSI-free beamforming, where the network can direct beams based on contextual data, bypassing the need for explicit channel estimation and feedback. By jointly processing these different data modalities, FNO can resolve ambiguities and correct errors that would be impossible to fix using channel data alone.
\end{itemize}

%
\subsection{FNO as a Nonlinear Inverse Operator}
\begin{itemize}
\item \textbf{Signal Recovery:} Practical MIMO hardware systems, such as power amplifiers, exhibit nonlinear distortions, especially at high frequencies. These nonlinear effects severely disrupt signal orthogonality, causing a sharp decline in the performance of traditional linear detectors.

\item \textbf{FNO Solution:} FNO can learn a powerful nonlinear inverse operator, which treats the nonlinear distortions across the entire transceiver chain as a complex, unknown nonlinear operator and then learns its inverse process. The input is the distorted received signal function, and the output is the recovered original symbol sequence. Through end-to-end training, FNO can implicitly model and compensate for various hardware nonlinearities, serving as a unified and robust signal recovery module.
\end{itemize}
\subsection{FNO as a Functional Autoencoder}
\begin{itemize}
\item \textbf{End-to-End Transceiver:} The traditional modular design of transmitter-channel-receiver suffers from information bottlenecks, where the local optimization of each module does not guarantee end-to-end global optimality.

\item \textbf{FNO Solution:} The communication transceiver can be viewed as a functional autoencoder operating in continuous function space that represents signals and information. The transmitter-side FNO (encoder) learns to map a source information function to a waveform function suitable for propagation. The receiver-side FNO (decoder) learns to recover the source information from the received signal function. By jointly training to optimize an end-to-end performance metric, FNO can discover superior communication strategies that transcend the limitations of modular design.
\end{itemize}

\subsection{FNO as a Semantic Codec}
\begin{itemize}
\item \textbf{Semantic Communications:} Building on the foundation of bit information flow-based transceivers, the semantic communication technology further introduces a new dimension for communication system design, which aims to transmit meaning beyond mere bits \cite{11175642}. 

\item \textbf{FNO Solution:} FNO can serve as a semantic codec in function spaces. An encoder FNO learns to extract a continuous semantic feature function from source data, e.g., images or videos.  A decoder FNO then reconstructs the semantic information from the received, and possibly distorted, feature function. Operating in function spaces allows FNO to learn more abstract and robust semantic representations that are effectively resilient to channel noise and fading.
\end{itemize}

\section{Case Studies and Numerical Results}
To concretely demonstrate the practical advantages of FNO, we explore its application in channel modeling and estimation for two challenging MIMO architectures.

\subsection{Channel Modeling for Holographic MIMO Systems}
\begin{figure*}[t]
\centering
\includegraphics[width=6.5in]{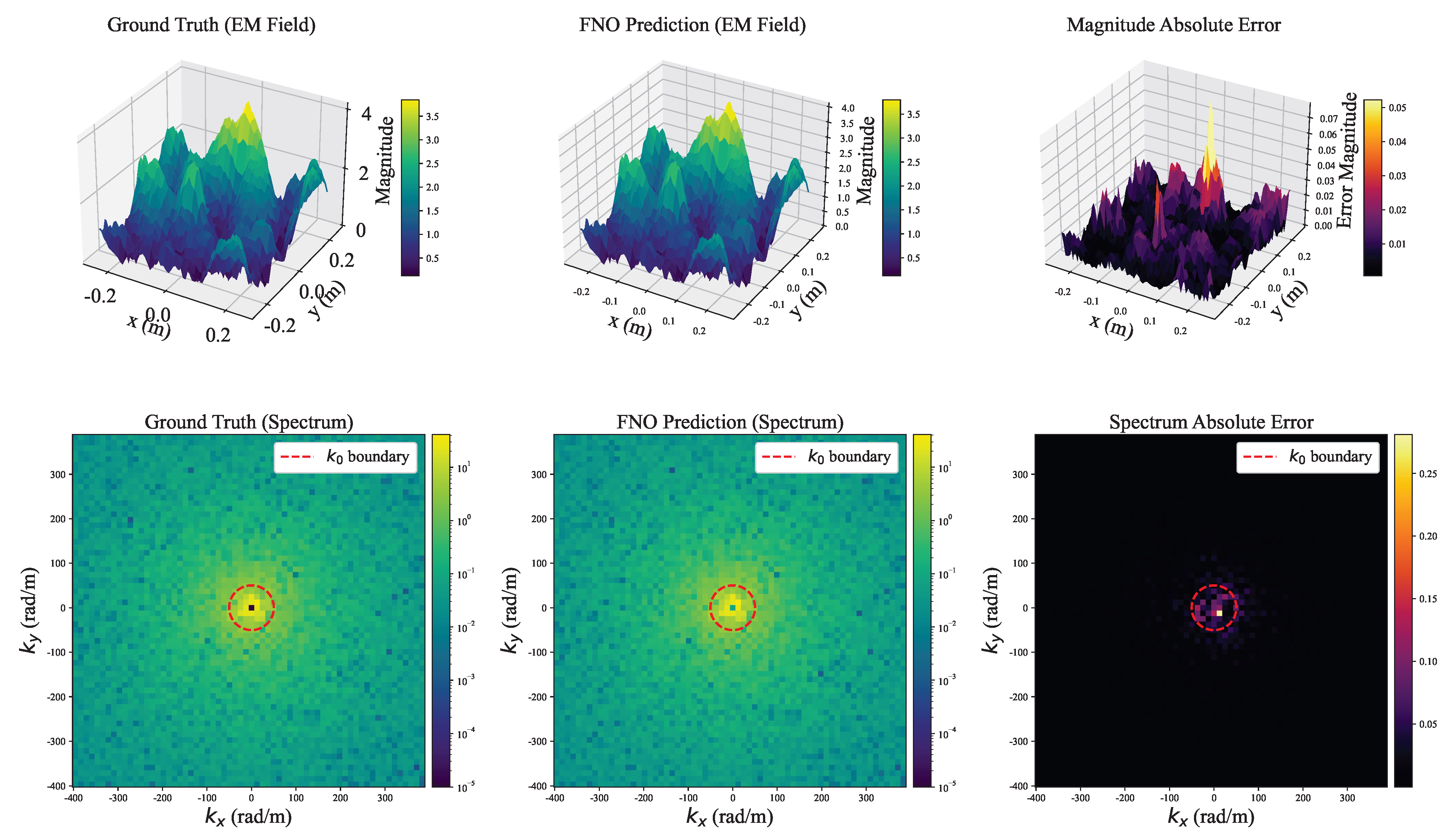}
\caption{Channel modeling performance for holographic MIMO systems. The top row compares the magnitude of EM field across a 2D spatial plane, showing that the FNO prediction (center) closely matches the ground truth (left) with minimal absolute error (right). The bottom row analyzes the frequency spectrum, where the FNO prediction successfully replicates the key physical property of the ground truth by concentrating its energy within the wavenumber $\kappa_0$ boundary.
}
\label{fig4}
\end{figure*}
Holographic MIMO treats the antenna array as a quasi-continuous EM surface, in which the channel is no longer a discrete matrix but an integral operator that maps a continuous current distribution function \( \mathbf{J}(s) \) on the transmit aperture to a continuous EM field function \( \mathbf{E}(r) \) on the receive plane. To address this problem, we employ FNO to learn this function-to-function channel operator \( G: \mathbf{J}(s) \rightarrow \mathbf{E}(r) \), using it as an efficient surrogate model.


%
%
%

As illustrated in Fig.~\ref{fig4}, the top-row 3D surface plots of the EM field show a high degree of visual similarity between the FNO prediction and the ground truth, with the magnitude of the absolute error being minimal. The generation time per channel instance is only 2.04 ms on the Intel i9-12900K CPU. This indicates that FNO can accurately capture the complex spatial interference patterns, and implicitly learn the physical propagation laws. Furthermore, the bottom-row analysis in the frequency domain reveals the spectrum of the ground truth EM field is concentrated within a circle defined by the wavenumber \( k_0 \). This is consistent with the physics of propagating waves, as spatial frequencies higher than \( k_0 \) correspond to evanescent waves that decay exponentially and do not reach the far field. The predicted spectrum of FNO almost perfectly replicates this behavior, concentrating its energy within the \( k_0 \) boundary and correctly suppressing the high-frequency components.

\subsection{Channel Estimation for Flexible Intelligent Metasurfaces}

The wireless channels in FIM are directly coupled to its physical deformation shape, where the channel $\mathbf{{h}}$ is a function of the high-dimensional deformation vector \( \zeta \). Since the FIM geometry can vary continuously, the space of possible channel states is infinite. The core challenge is to design a procedure that takes a limited set of noisy channel measurements from pre-defined pilot shapes $\{\mathbf{\hat{h}}(\zeta_{m})\}_{m=1}^{M}$, and produces an accurate estimate $\mathbf{\hat{h}}(\zeta_\text{target})$ for any arbitrary target deformation $\zeta_\text{target}$ with limited pilot overhead $M$ \cite{xiao2025channel}. The proposed FNO reframes the estimation problem as learning a continuous and non-linear operator that maps the set of pilot measurements and a query for a target deformation to the channel.

\begin{figure}[!t]
\centering
\includegraphics[width=3.5in]{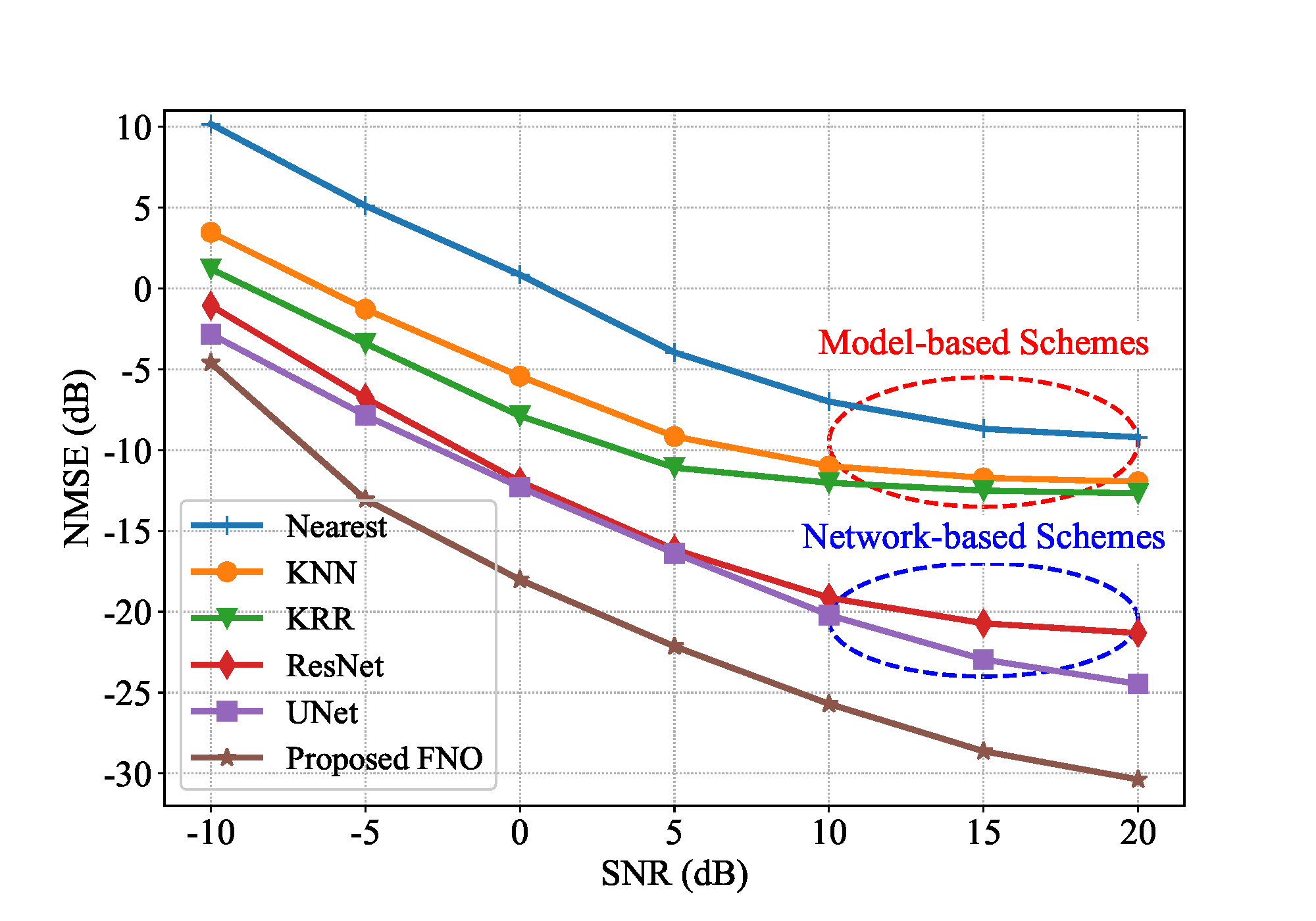}
\caption{Channel estimation for FIM systems, where we consider the model-based schemes, i.e., Nearest Interpolation, K-Nearest Neighbors (KNN) and Kernel Ridge Regression (KRR), and typical neural network-based schemes, i.e., ResNet and UNet.}
\label{fig5}
\end{figure}

Fig. 5 presents that FNO consistently achieves a lower normalized mean squared error (NMSE) than model-based and network-based schemes across the entire SNR range. This superior performance stems from the fact that while traditional methods rely on fixed model assumptions and conventional neural networks act as generic function approximators, the FNO is specifically designed to learn the complex, underlying physical relationship between the FIM geometry and the resulting channel response. Its operator-learning paradigm, which maps between continuous function spaces, and its use of global convolutions in the Fourier domain provide a strong inductive bias for wave-based physics. This allows FNO to learn a more accurate and generalizable model of the channel operator. Consequently, FNO demonstrates a more powerful generalization capability, enabling it to accurately predict the channel state for unseen FIM shapes with fewer pilot probes than conventional networks. 

\section{Open Challenges and Future Directions}
While FNO holds immense promise, its practical application is not without hurdles. Several key challenges need to be addressed to unlock its full potential.

\subsection{High-Quality Data Acquisition}

Similar to all data-driven methods, the performance of FNO is contingent on the availability of high-quality training data, which poses significant challenges in the wireless domain. {Unlike traditional AI tasks that require statistically representative discrete samples, training an FNO as a physics surrogate requires data that accurately samples an infinite-dimensional function space. For dynamic systems, such as flexible MIMO, this necessitates generating datasets covering a vast range of continuous physical geometries, a task that is extremely costly and time-consuming, whether through high-fidelity EM simulations or extensive real-world measurements.}

\subsection{Modeling Hardware and Physical Constraints}
Real-world hardware has physical limitations that must be incorporated into the FNO-based physical-layer design. The output of an FNO is typically a continuous function. However, hardware controls are often discrete, e.g., the phase shifts of an metasurface. This necessitates efficient and intelligent quantization of the FNO output. Moreover, the movement of flexible antennas is constrained by speed and acceleration limits. These mechanical constraints must be integrated into the FNO-based control logic to ensure feasibility.

\subsection{Closed-Loop Control and Robustness}


Deploying FNO in a live system requires closed-loop control mechanisms \cite{11016699}, {the challenge transcends typical AI prediction. For future reconfigurable communication systems, the FNO may act as an actuator where its output, e.g., an optimal surface shape, physically alters the system, which in turn influences the next input of FNO. This creates a tight physical feedback loop that must account for dynamic constraints, e.g., mechanical movement speeds, and the continuous-to-discrete gap between the functional output of FNO and the discrete control signals of hardwares. This level of physical interaction is a more complex control problem than that faced by standard and open-loop AI predictors.}


\subsection{Real-Time Implementation and Hardware Acceleration}
For practical deployment, FNO-based models must meet the stringent latency requirements of wireless systems. While FFT is efficient, the overall complexity of a deep FNO model can still be significant. Research into model compression, quantization, and dedicated hardware accelerators for FNO is needed to enable real-time inference on base station hardware. In particular, knowledge distillation techniques could be used to train smaller and faster student FNOs that mimic the performance of a larger, more complex teacher model.


\section{Conclusion}
Next-generation MIMO systems are set to redefine the limits of wireless communications, but they bring a host of new challenges for physical layer signal processing. FNO, with its foundation in functional analysis and its efficient Fourier-based architecture, presents a compelling solution. By learning the fundamental operators that govern wave propagation and channel responses, FNO can provide highly accurate, computationally efficient, and generalizable solutions for various signal processing tasks. In this article, through case studies in holographic and flexible MIMO systems, we have shown that FNO can outperform both traditional model-based methods and conventional deep learning architectures. While challenges remain, the continued development and integration of FNO into the wireless communication stack will be a key enabler for the high-performance, intelligent, and adaptive 6G networks.

\ifCLASSOPTIONcaptionsoff
  \newpage
\fi

\bibliographystyle{IEEEtran}
\bibliography{IEEEabrv,refs.bib}
\end{document}